\documentclass[amssymb,prb,aps,twocolumn,showpacs]{revtex4}
\usepackage{graphicx}
\begin{document}
\title{Electronic structure of noble metal impurities in
semiconductors: Cu in GaP}
\author{O.V. Farberovich$^{1}$, A.
Yaresko$^{2}$, K. Kikoin$^{3}$ and V. Fleurov$^{3}$}
    \affiliation{$^1$ Department of Physics, Ben-Gurion University,
    Beer-Sheva 84105, Israel.\\
    $^2$Max-Planck-Institut f\"ur Physik Komplexer Systeme,
    N\"othnitzer Str. 38, D-01187 Dresden, Germany.\\
    $^3$School of Physics and Astronomy, Beverly and Raymond Sackler
    Faculty of Exact Sciences, Tel Aviv University, Tel Aviv 69978,
Israel.}
\date{\today}
\begin{abstract}

A numerical method for calculation of the electronic structure of
transition metal impurities in semiconductors based on the Green
function technique is developed. The electronic structure of 3$d$
impurity is calculated within the LDA+U version of density
functional method, whereas the host electron Green function is
calculated by using the linearized augmented plane wave
expansion. The method is applied to the Cu impurity in GaP. The
results of calculations are compared with those obtained within
the supercell LDA procedure. It is shown that in the Green
function approach Cu impurity has an unfilled 3d shell. This
result paves a way to explanation of the magnetic order in dilute
Ga$_{1-x}$Cu$_x$P alloys.

\end{abstract} \maketitle

\section{Introduction}

The experimental and theoretical studies of dilute magnetic
semiconductors (see, e.g. the recent review \cite{McD}) have
revived the interest to details of reconstruction of the
electronic structure of host materials induced by transition metal
ions and concomitant defects. This interest stems from the
fact that the simple Vonsovskii-Zener model of $s$-$d$ exchange is
apparently not sufficient for an exhaustive explanation of  the
behavior of the most popular system (Ga,Mn)As,\cite{Bouz1} not to
mention the wide-gap materials like (Ga,Mn)N, (Zn,Co)O,
(Ti,Co)O$_2$.\cite{Stras,Coey,Grif} Not only the localized spin of
magnetic ions but also the acceptor or donor-like states in the
energy gap related to these ions are involved in the indirect exchange between the magnetic ions responsible for the long-range magnetic order. The nature of these states is the matter of a vivid discussion in the current literature.

In particular, an isolated Mn impurity in GaAs creates a 0.11 eV
acceptor level relative to the top of valence band. Besides, the
electrons in the half-filled $3d$ shell form resonance levels in the
middle of this band because of an anomalous stability of the
half-filled 3d$^5$ shell.\cite{Zung86a,KF94} Since the
substitution impurity Mn$^{2+}(3d^5)$ is negatively charged
relative to the host semiconductor, localizing a hole makes this
defect neutral, and the binding energy of this hole is provided by
the combined action of the Coulomb potential, central cell
substitution potential, hybridization and, maybe, $s$-$d$
exchange.\cite{BhaBe,Krsta} At a high enough Mn concentration, these
acceptor levels form an impurity band and eventually merge with the
hole states near the top of the valence band (see Ref. \onlinecite{merge} for a
detailed discussion of the current experimental situation).

According to the available calculations of the electronic spectra of
an isolated Cu in GaP,\cite{Zung85} the copper impurity should
have a similar electronic structure. Due to the special stability
of the filled 3d$^{10}$ shell all the 3d levels of the Cu impurity
are expected to be occupied in the ground state, and the electrical
neutrality of Cu impurity should be ensured by capturing two holes
on Cu-related acceptor levels close to the top of the valence
band, so that the resulting electron configuration can be denoted
as Cu(d$^{10}\bar p^2$). Indeed such acceptor states were found in
CaP:Cu samples,\cite{Ledebo} although at that time the nature of
these states remained unclear.

Recently, ferromagnetism with a high Curie temperature in $p$-type
Cu-doped GaP was detected.\cite{Gupta} The EPR signal of the Cu$^{2+}$
state indicates that the 3$d$ shells of Cu impurities are unfilled
in this material in contradiction to the results of previous
numerical calculations. This discrepancy gives us a motivation to
revisit the problem.

We present in this paper the results of numerical calculations of
the electronic structure of Cu-doped GaP. Two different
computation schemes are used, which give mutually complementary
information about the behavior of weakly and strongly doped
materials. The first one is the conventional local density
approximation (LDA) scheme applied to the lattice of
Cu$_{x}$Ga$_{1-x}$P supercells. Similar methods were used for
Mn$_{x}$Ga$_{1-x}$Pn materials with Pn=As,N,P.\cite{Shik,Kronik}
The second method is based on the local Green function
approach.\cite{FK86} In this method the hybridization between the
local impurity $d$-orbitals and Bloch waves in the host
semiconductor is calculated exactly, without any kind of
artificial periodic boundary conditions, and approximations are
made only when taking into account the short-range part of
substitution impurity potential.

\section{Green function approach for isolated impurity}

A Green function calculation procedure based on the microscopic
Anderson model \cite{And61} was proposed three decades ago
\cite{Fleurov1,Haldane} and later on summarized in Ref.
\onlinecite{KF94}. This procedure deals with the local Green
function
\begin{equation}\label{green}
G_{\rm imp}({\bf r,r'},z) = \sum_\lambda |\lambda\rangle\langle
\lambda|(z - H)^{-1}|\lambda\rangle\langle \lambda|~.
\end{equation}
The set $|\lambda\rangle$ includes both the electron states
$\phi^\sigma_{i_a}(\bf r)$ of the electrons localized in the
d-shell of impurity atom and the states $\psi_{b,\gamma\mu\sigma}(\bf r)$, which stand for "the Bloch tail" of the impurity wave function. These states describe the distortion inserted by a substitution impurity in the spectrum of
a host crystal. They are superpositions of the Bloch waves,
$\psi_{b,\gamma\mu\sigma} = \sum_{{\bf k}n}C^{\gamma\mu}_{{\bf
k}n}\psi_{{\bf k}n\sigma}$, where ${\bf k}$ and $n$ are the wave
vector and the band index respectively, $\sigma$ is the spin
quantum number. Here $\gamma$ is the index of the irreducible
representation of the point group characterizing the symmetry of
impurity and its surrounding, and $\mu$ denotes its row. Therefore
the function $G_{\rm imp}$ is diagonal in $\gamma\mu$
representation. The full Hamiltonian $H$ includes the kinetic and
potential energies of all electrons in the impurity atom and in
the host crystal, as well as the Coulomb and exchange interactions
between these electrons. The projection procedure (\ref{green}) is
exact in principle, and the poles of the Green function $G_{\rm
imp}$ describe both continuous and localized impurity related
states in the doped crystal. In the practical realization of this
method some approximations are unavoidable. The main
simplification, which we use here is the approximate treatment of
the substitution potential
$$
\Delta V({\bf r - R}_0)= V_{\rm eff}({\bf r-R}_0)- V_h^0({\bf r -
R}_0),
$$
where $V_h^0({\bf r-R}_0)$ is the potential landscape for an
electron in the host gallium atom in the site ${\bf R}_0$ and
$V_{\rm eff}({\bf r-R}_0)$ is the self-consistent potential for
the electrons in the $3d$ shell of the Cu ion substituting for Ga
in this site (see Section III for detailed definition of these
potentials). We suppose that this potential is localized within
the defect cell of the doped crystal. The "local substitution
potential" approximation influences only the description of
$p$-type acceptor states in the lower part of the forbidden energy
band. It ignores possible contribution of the Coulomb component of
substitution impurity potential. This contribution is known to be
small in the case of (Mn,Ga)As,\cite{merge} and one may hope for a
similar situation in (Cu,Ga)P. The principal advantage of the
local substitution potential is that in this case the system of
Dyson equations for the impurity-related components of the Green
function (\ref{green}) defined as
\begin{equation}\label{greenin}
G_{\gamma\mu}(z) = \langle\gamma\mu|(z - H)^{-1}|\gamma\mu\rangle
\end{equation}
may be solved analytically \cite{FK86}. It yields the
equation
\begin{eqnarray}\label{dgreen}
G^{-1}_{\gamma\mu}(z) = z - \varepsilon_{d\gamma} - {\cal
M}_{\gamma}(z)/Q(z),
\end{eqnarray}
for the $d$-electron Green function. The positions of electron
$d$-levels $\varepsilon_{d\gamma}$ are found self-consistently as
a solution of the Schr\"odinger equation for Cu-related orbitals
in the crystalline environment. The self energy in the right-hand
side of Eq. (\ref{dgreen}) contains two contributions. The term
${\cal M}_{\gamma}(z)$ describes the hybridization between
the $d$-orbitals and the band electrons
\begin{equation}\label{massoper}
{\cal M}_{\gamma}(z)=\sum_{n{\bf k}}\frac{|M_{\gamma,n{\bf
k}}|^2}{z - \varepsilon_{n{\bf k}}}
\end{equation}
where the hybridization integral is
\begin{equation}\label{hyb}
M_{\gamma,n{\bf k}}= \int \psi^*_{d\gamma\mu}({\bf r})\Delta
V({\bf r})\psi_{n{\bf k}}({\bf r})d{\bf r}.
\end{equation}
The energy bands $\varepsilon_{n{\bf k}}$ and Bloch functions $
\psi_{n{\bf k}}({\bf r}) $ of the host GaP crystal are calculated
by means of the first principle full potential LAPW method
\cite{MaH,FaVa} (see Section III for details).

The factor
\begin{equation}\label{Kosla}
Q(z) = 1 - \Delta V_0 G_h^0(z).
\end{equation}
in Eq. (\ref{dgreen}) describes the short-range potential scattering,
where
\begin{equation}\label{sub}
\Delta V_0 = \sum_{n{\bf k}n{\bf k}'}\int \psi^*_{n{\bf k}}\Delta
V({\bf r - R}_0) \psi_{n{\bf k}'}d{\bf r}
\end{equation}
is the substitution impurity potential localized in the defect
shell,
\begin{equation}\label{greenol}
G_h^0(z) = \sum_{n{\bf k}}\langle n{\bf
k}|(z - H_0)^{-1}|n{\bf k}\rangle = \sum_{n{\bf k}}
\frac{1}{z - \varepsilon_{n{\bf k}}}
\end{equation}
is the single-site lattice Green function for the electrons in the
non-doped host crystal described by the Hamiltonian $H_0$.

As was shown in Ref. \onlinecite{FK86}, the Green function
(\ref{dgreen}) describes the hybridization between the impurity
$d$-electron orbitals and the electrons in
the imperfect host crystal, where the band electrons are
influenced by the potential scattering $\Delta V$. If this
scattering is strong enough, it results in splitting off of localized
levels from the top of the valence band. This effect is also taken
into account in (\ref{dgreen}): the positions of the corresponding
levels before the hybridization are determined by zeros of the
function $Q(z)$ in the energy gap of the host crystal.

One of the fundamental statements of the theory of transition
metal impurities in semiconductors \cite{Zung86a,KF94} is the
necessity to discriminate between the impurity levels in the gap
obtained as solutions of a self-consistent mean-field Schr\"odinger
equation for a doped crystal and the true addition/extraction
energy of a $d$-electron to/from the valence/conduction band. The
latter energies are determined by the energy balance of "Allen
reactions"\cite{Zung86a,KF94,Allen}
\begin{eqnarray}\label{allen}
\varepsilon^{n/n-1} = E(d^n)-E(d^{n-1}) - \varepsilon_v \nonumber \\
\varepsilon^{n+1/n} = \varepsilon_c - E(d^{n+1}) + E(d^{n})
\end{eqnarray}
Here $E(d^p)$ is the total energy of doped crystal with the impurity
having $p$ electrons in 3d shell.  Two Allen reactions describe
the electron transition from the top of the valence band
$\varepsilon_v$ to the empty neutral (acceptor) level and the
electron transition from an occupied charged (donor) level to the
bottom of the conduction band $\varepsilon_c$, so that the
energies (\ref{allen}) characterize the true positions of the
impurity levels with respect to the band edges in the presence of
strong Coulomb and exchange interactions. These energies do not
necessarily coincide with the mean-field solutions of the Schr\"odinger
equation due to the violation of Koopmans' theorem for the
impurity ions.

To minimize the mismatch between the single-electron and
many-electron states, Slater proposed a concept of "transition
state". According to his arguments, the ionization energy for a
state with $n$ electrons in the 3d shell, $E(d^n) - E(d^{n-1})$
may be approximated by the energy $\epsilon(n - 1/2)$ calculated
within the LDA single-electron calculation scheme. More refined
LDA+U approach \cite{Aza,Solov} takes non-Koopmans' corrections to
the single-electron energies into account explicitly (although still
semi-phenomenologically). In terms of the Allen energies
(\ref{allen}), the energy $U$ is just the difference between
$\varepsilon^{n+1/n}$ and $\varepsilon^{n/n-1}$. We test below
both the LDA+U method of Green function calculations and the standard LDA
supercell description of dilute (Gu,Ga)P semiconductor.

\section{Impurity Green functions in LDA+U approximation}

To realize a numerical version of the Green function method we use
the local density approximation (LDA) and its modification LDA+U,
which takes into account strong electron-electron correlations on
the impurity site. This section outlines the application of the
LDA+U method to systems with local defects with a particular
emphasis on the transition metal impurities for which the resonant
scattering in the d ($l=2$) channel plays a crucial role. Here we
present only the principal features of the scheme leaving many
more mathematical details and definitions in Appendix. In this
section we retain the spin index, having in mind to use the
spin-unrestricted LSDA+U version of this method for the calculation of
spin properties of dilute magnetic semiconductors, although in the
practical calculations below only the spin-independent LDA+U
version is used.

The LDA + U method incorporates a correction to the LDA energy
functional which provides an improved description of the electron
correlations. The principal idea of the LDA + U method is to
separate the electron system into two subsystems of the localized
$d$-electrons for which the Coulomb interaction is accounted for
by the Hubbard repulsion term $\frac{1}{2}U \sum_{m\neq m'}\rho_m
\rho_{m'}$ in the model Hamiltonian whereas the delocalized $s$-
and $p$-electrons are described by an orbital independent one
electron potential $V^{LDA}({\bf r})$.

As a result the  impurity Green function (\ref{green}) is defined
by the Dyson equation
\begin{equation}\label{dyson}
G_{i}^\sigma(z) = G_{i}^{(0)\sigma}(z) \left[1 + {\cal
 \widetilde M}_{i}^\sigma(z) G_{i}^\sigma(z) \right]
\end{equation}
where
\begin{equation}\label{selfmod}
 \widetilde{{\cal M}}_{i}^\sigma(z) = {\cal
M}^\sigma_{i}(z)/Q(z).
\end{equation}
The Green function (\ref{dgreen}) is a solution of this equation.
We work in the spherically symmetric local basis $(i\equiv plm)$
instead of cubic harmonics expansion $(i\equiv p\gamma\mu)$ used
in (\ref{dgreen}). Here $p$ is the index of repeating irreducible
representations $\gamma\mu$ or $lm$, the analog of the principle
quantum number $n$ in a spherical atom.

The bare $d$-electron Green function
\begin{equation}\label{LDA+U}
G_{i}^{(0)\sigma}(z) = \frac{1}{z - \varepsilon_{i} - \Delta
{V^\sigma_{ii}}^{LDA+U}}
\end{equation}
includes intraatomic correlations in the form of LDA+U potential
consisting of three terms,
\begin{equation}\label{dev}
V\sigma_{ii,\sigma}^{LDA+U}=\Delta V_{ii,\sigma}^{LDA}
 + \Delta V_{ii,\sigma}^{U} + \Delta V_{ii,\sigma}^{dc}.
\end{equation}
Here  the first term is the substitution LDA potential
\begin{eqnarray}\label{dev1}
&&\Delta V_{plm,plm;\sigma}^{LDA} =
\nonumber \\
&& \sum_{l''} G_{lm,L''0}^{lm}  \int_0^{r_{emb}} dr r^2 \Delta
V_{l''0,\sigma }^{LDA} (r) R_{pl\sigma}^2(r)~,
\end{eqnarray}
The second term is the electron-electron interaction potential
in the 3d shell,
\begin{widetext}
\begin{equation}\label{dev2}
\Delta V_{ii,\sigma}^{U} = \sum_{m''}\left[\left(U_{mmm''m''} -
U_{mm''m''m}\right) \rho^{-\sigma}_{plm'',plm''} +
U_{mmm''m''}\rho^{\sigma}_{plm'',plm''}\right]
\end{equation}
\end{widetext}
and the last term is the double counting compensation potential,
parametrized as
\begin{equation}\label{dev3}
\Delta V_{ii,\sigma}^{dc} = -U(\sum_{m\sigma}n^{}_{plm,\sigma} -
\frac{1}{2}) + J(\sum_m n_{plm,\sigma} - \frac{1}{2}).
\end{equation}
Here we introduced the occupational matrix
$$
\rho^{\sigma,pl}_{mm'} = -\frac{1}{\pi} \mbox{Im}
\int_{\varepsilon_b}^{\varepsilon_F}
\left[G(z)\right]_{mm'}^{\sigma,pl} dz
$$
as a contour integral of the relevant matrix elements of the LDA+U Green function
(\ref{LDA+U}). The Slater integrals\cite{Czy} in the atomic limit read
\begin{eqnarray}\label{slat}
&&U_{m_1m_2m_3m_4} \equiv \langle m_1,m_3|V^{ee}|m_2,m_4\rangle
\nonumber
\\
&& = \sum_{k=0}^{2l} a^k_{m_1m_2m_3m_4} F^k(l,l)
\end{eqnarray}
where the coefficients
$$
a^k_{m_1m_2m_3m_4} =
$$$$
\frac{4\pi}{2k+1} \sum_{n = -k}^k \langle lm_1|Y_{kn}|lm_2\rangle
\langle lm_3|Y^*_{kn}|lm_4\rangle
$$
can be expressed in terms of the Gaunt coefficients
$G_{lm,l'm'}^{l^{\prime\prime}m+m'}$ (see Appendix A.3).

The hybridization matrix elements (\ref{hyb}) in the numerator of
the mass operator now take the form
\begin{equation}\label{matelem}
M^{\sigma}_{n{\bf k}, i} =
$$$$
\int_{\Omega_{emb}}
{\phi^{\sigma*}_{i}}({\bf r}) \Delta V({\bf r}) \Psi^{LAPW}_{n{\bf
k}}({\bf r} - \mbox{\boldmath$\tau$}_s)\Theta({\bf r} -
\mbox{\boldmath$\tau$}_s) d{\bf r}.
\end{equation}
Here the Bloch wave functions $\Psi^{LAPW}_{n{\bf k}}$ are
calculated by means of the linearized augmented plane wave (LAPW)
method, $\mbox{\boldmath$\tau$}_s$ is the vector connecting
substitution impurity site taken as the point of origin with its
nearest P neighbors in the zinc-blend lattice.
\begin{figure}[tbp!]
\begin{center}
\includegraphics[width=0.8\columnwidth]{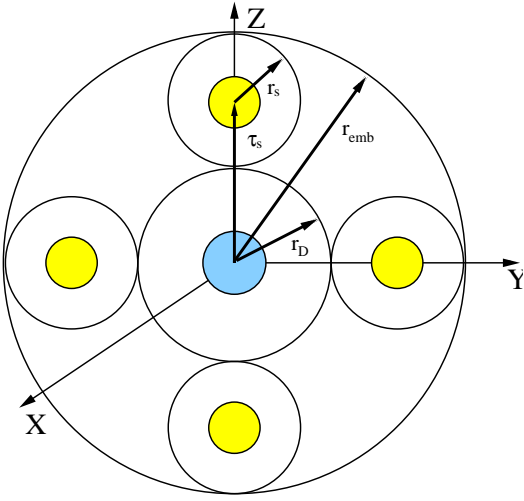}
\end{center}
\caption{(Color online) The embedded sphere and coordinate system
used in our calculations.} \label{spheres}
\end{figure}

In the impurity version of FLAPW method the defect site occupied
by a Cu ion is surrounded by the "embedded sphere" with the radius
$r_{\rm emb}$ which includes the impurity sphere with the radius
$r_D$ (muffin-tin region, where the impurity potential is
non-zero). Muffin-tin spheres $r_s = (r_{emb} - r_D)/2$ with a
non-zero host lattice potential surround also the neighboring Ga
sites (see Fig. \ref{spheres}). The impurity centered basis set is
chosen as a set of the linear augmented spherical wave (LASW)
functions [see Eqs. (\ref{wf_1}), (\ref{radial})]. In  accordance
with the LASW method, a set of Bessel functions is used in the
remaining part of the sphere $r_{\rm emb}$. The wave functions in
the two regions are matched by the standard boundary conditions
imposed on the wave function and its derivative. The Bloch
functions $\Psi^{FLAPW}_{n{\bf k}}({\bf r})$ of the host GaP
crystal outside the embedded sphere are obtained by the
self-consistent FLAPW method. Using the impurity centered local
LASW functions we calculate the matrix elements of the host Green
function projected onto the local spin polarized LAPW functions in
the spherical interstitial site. After matching the boundary
conditions (see Appendix), the matrix element (\ref{matelem}) is
transformed into
\begin{widetext}
\begin{equation}\label{matelspher}
M^{\sigma}_{n{\bf k}, i} = \frac{4\pi\tau_s^2}{\sqrt{\Omega_0}}
\sum_{\varrho=1}^{N} v_n({\bf k}_\varrho) \sum_{L''M''}\sum_{lm} i^l
Y^*_{lm}(\widehat{\bf k}_\varrho) G^{lm}_{lm-M'', L'' M''}
\int_0^{r_{emb}} dr r^2 R_{pl}^\sigma (r) \Delta
V_{L''M''}(r){\Phi_l^s(k_\varrho,r)}^{LAPW}
\end{equation}
\end{widetext}
($\varrho$ stands for the vectors of reciprocal lattice, see Appendix).
As was mentioned above, substituting Ga for a Cu impurity results
also in an appearance of a potential component of the impurity
potential, which is taken into account approximately by adopting
the Koster-Slater-like single site scattering
approximation.\cite{FK86} Then in accordance with Eq.
(\ref{dyson}), one may introduce the modified mass operator
${\cal\widetilde M}_{i_a}^\sigma$ (\ref{selfmod}), where the zeros
of the operator $Q(z)$ (\ref{Kosla}) determine the impurity
states, which arise in the doped crystal due to the potential
scattering only.

The scattering amplitude $\Delta V_0$  is calculated by substituting the LAPW wave functions ${\Psi_{n'{\bf k}'}}^{LAPW}({\bf r})$ for the Bloch functions in Eq. (\ref{sub}).

%
%\begin{equation}\label{potscat}
%\Delta V_0 = \sum_{n{\bf k},n'{\bf k}'} \Delta V_{{n{\bf
%k},n'{\bf k}'}}
%\end{equation}
%
%with
%
%$$
%\Delta V_{{n{k},n'{\bf k}'}} = \int_{\Omega_{emb}}
%{\Psi^*_{n{\bf k}}}^{LAPW}({\bf r}) \Delta V({\bf r}) {\Psi_{n'{\bf
%k}'}}^{LAPW}({\bf r})d{\bf r}.
%$$

As a result the equation for the deep level energy determined as a
pole of the impurity Green function (\ref{dgreen}) within the
framework of the LDA+U technique reads
\begin{equation}\label{resontscat}
z - \varepsilon_{i} - \Delta V^{LDA+U}_{ii} = \widetilde{ {\cal
M}}_{i}^\sigma(z).
\end{equation}
It takes into account the resonance part of the scattering
amplitude in the d ($l=2$) channel and its mixing with the potential
scattering states arising in the p ($l=1$) channel. \cite{FK86}

The adspace augmentation \cite{adspace} is used to represent the
Green function (or resolvent) $G(z)$ for the GaP crystal with a Cu
impurity in the matrix form, Eq. (\ref{augmgreen}). The impurity
augmented Green function is subdivided into two blocks, of which
the upper left corner block $G^0_A(z)$ is constructed using the
basis of $i$ orbitals where $i$ refer to the $i$-th state with the
energy $\varepsilon_{i}$ of the isolated adatom. The host is
represented by the lower right corner block $G_h^0(z)$.

It is worth emphasizing that such a direct introduction of the new
adatom related states is very effective in the matrix formulation.
Since the high energy part of the spectrum of the differential
operator is well suited for the description of the strongly
localized $d$-type impurity states,\cite{Lind} the issue of the
necessary number of the host crystal bands becomes crucially
important. The direct introduction of the $d$-states drastically
simplifies the problem. The Dyson equation may be then split into
two independently solvable equations (see Appendix) which finally
allows one to carry out the calculations of the GaP host Green
$G^0_h(z)$ using only 15 bands.

The problem is treated self consistently, starting with the trial
set of LAPW functions obtained with the help of the impurity
potential, which in the zero's approximation is just a sum of the
atomic potentials of the defect crystal. The self-consistency
procedure for $\Delta V({\bf r})$ is carried out in a mixed
fashion. The first two iterations use the arithmetic average
scheme, which later on is effectively substituted by the Aitken
scheme.\cite{Aitken} Just seven iterations produce the $\approx
2\cdot 10^{-4}$ Ry self-consistency.

The equations presented in this Section will be our working
formulas for the LDA+U calculations of the Ga(Cu)P compound where
the Cu atoms substituting Ga host atoms will be considered as
isolated impurities. A possible exchange interaction between the
Cu atoms and the resulting magnetic effects will be considered
elsewhere.

\section{Discussion of the results}

This section presents the results of calculation of the electronic
structure of Cu$_x$Ga$_{1-x}$P obtained by means of the two
methods, both using the LDA approximation. The Green function
approach is based on the band structure calculated by means of the
FLAPW method discussed in the previous section. The supercell
approach uses the AS-LMTO method\cite{Wil} for the band
calculations. The Vosko\cite{Cap} and Perdew-Wang \cite{Per}
parametrization scheme is used for the calculation of the
exchange-correlation potential in the former and latter
approaches, respectively. Brillouin zone (BZ) integration is
performed using the improved tetrahedron method.\cite{tet}

\begin{figure}[tbp!]
\begin{center}
\includegraphics[width=0.95\columnwidth]{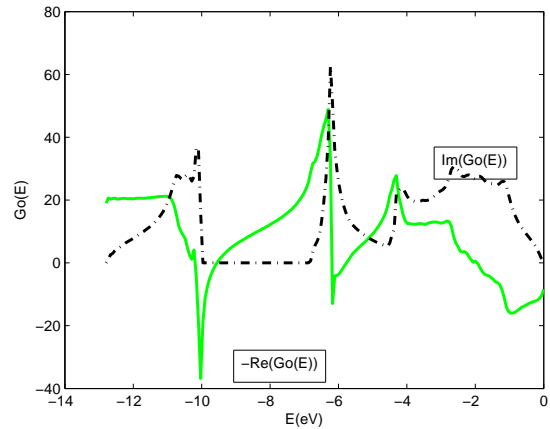}
\end{center}
\caption{(Color online)The functions Re$(G_h^0)$ and Im$(G_h^0)$
for GaP.} \label{Gh}
\end{figure}

According to the present FLAPW and LMTO calculations, the undoped
GaP is a semiconductor with the 1.83 eV FLAPW indirect gap and
1.61 eV ASA-LMTO gap between the top of the valence band (VB) at
the $\Gamma$ point and the bottom of the conduction band (CB) at
the (0,0,0.875) point close to the $X$ point of the fcc BZ. A
direct gap of 1.77 eV opens at the $\Gamma$ point. The 6.8 eV
width valence band is formed by the strongly hybridized P $p$ and
Ga $s$ and $p$ states while the states at the top of VB in the
vicinity of $\Gamma$ are formed by the P and Ga $p$ states with a
dominant contribution of the former. The band originating from the
P related $s$ states hybridized with the Ga related $s$ states is
found between $-$12.5 and $-$9.5 eV and separated by a gap of 2.8
eV from the bottom of the valence band. The density of states (DOS)
of GaP is visualized in Fig. \ref{Gh} as the imaginary part of the
Green function $G^0_h$ (\ref{greenol}) calculated by the LAPW method.

A similar picture is obtained by direct band structure calculations
within the ASA-LMTO method. The difference in the widths of the energy
gaps only weakly influences the structure of the Cu-related states in
the energy spectrum of doped samples. We start the discussion of these
states with a discussion of the supercell calculations.

\subsection{Supercell energy spectrum of ${\rm Cu_{1-x}Ga_xP}$}

The electronic structure of Cu$_x$Ga$_{1-x}$P with $x$ varying from
0.125 down to $\sim$0.016 was calculated using $2a\times2a\times2a$, $3a\times3a\times3a$, and $4a\times4a\times4a$ supercells of the cubic zinc-blend lattice. Calculations for $x$=0.125 (1/8), 0.063 (1/16), and 0.031 (1/32)
were performed for $F\overline{4}3m$ (216) fcc, $I\overline{4}3m$
bcc (217), and $P\overline{4}3m$ (215) simple cubic unit cells,
respectively. The face-centered cubic cells with $a=3a_0$ and
$a=4a_0$ allowed to simulate compositions with $x\approx0.037$
(1/27) and $x\approx0.016$ (1/64). In all the calculations the Ga
ion in the (0,0,0) position was substituted by the Cu ion with the
same atomic sphere radius. This way the tetrahedral ($T_d$)
symmetry of the Cu impurity site was preserved. The positions of
host atoms around the Cu impurity were not relaxed.

\begin{figure}[tbp!]
\begin{center}
\includegraphics[width = 0.95\columnwidth]{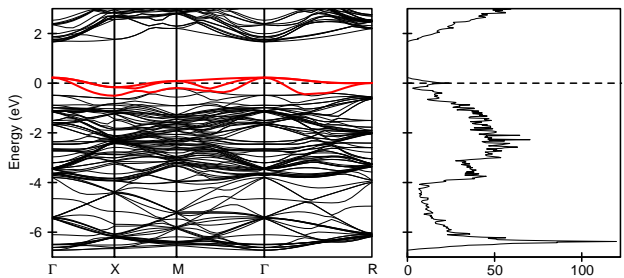}
\end{center}
\caption{(Color online) Bands calculated along some high symmetry
directions and the total DOS for Cu$_{x}$Ga$_{1-x}$P with
$x$=0.031.} \label{Fig:bnd32}
\end{figure}

Upon the Cu substitution Cu$_{x}$Ga$_{1-x}$P becomes a metal with each Cu impurity creating 2 holes in the valence band. At all the compositions $x$ studied in the present work the Fermi level ($\varepsilon_F$) crosses the three bands which are triply degenerate at the highest energy in the $\Gamma$ point. At $x$=0.063 the top of the valence band lies 0.42 eV above $\varepsilon_F$ and moves to 0.13 eV as the Cu concentration decreases to $x$=0.016. As an example, bands calculated along some high symmetry directions for Cu$_{x}$Ga$_{1-x}$P with $x$=0.031 are show in Fig.
\ref{Fig:bnd32}. At this Cu concentration the top of the valence band is
situated 0.22 eV above $\varepsilon_F$.

\begin{figure}[tbp!]
\begin{center}
\includegraphics[width=0.95\columnwidth]{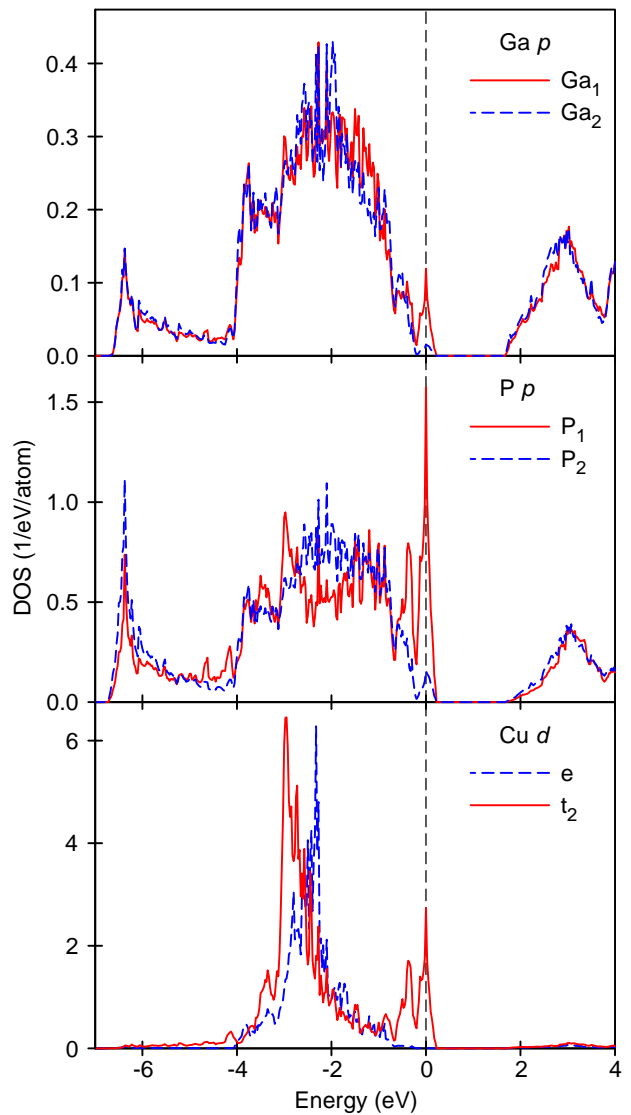}
\end{center}
\caption{(Color online) A symmetry resolved density of Cu $d$
states (lower panel) and the density of P $p$ (middle panel) and
Ga $p$ (upper panel) states calculated for Cu$_{x}$Ga$_{1-x}$P
with $x$=0.031.} \label{Fig:dos32}
\end{figure}

Figure \ref{Fig:dos32} (lower panel) shows the density of Cu $d$
states in Cu$_{x}$Ga$_{1 - x}$P with $x$ = 0.031 projected onto
the irreducible representations $e$ and $t_2$ of the $T_d$
symmetry group. The densities of $p$ states of the nearest (P$_1$
and Ga$_1$) and next nearest (P$_2$ and Ga$_2$) neighbors of the
Cu impurity are presented in the middle and upper panels of
Fig.~\ref{Fig:dos32}.

The calculations show that the Cu $d$ shell is almost completely
filled and the Cu valency is close to $1 +$. Cu $d$ states of $e$
symmetry ($3z^2 - 1$ and $x^2 - y^2$) form a DOS peak centered at
$ - $2.5 eV. They are completely occupied and do not contribute to
the bands crossing the Fermi level. The main peak of the density
of the $t_2$ ($xy$, $yz$, and $zx$) states is located at $ - 3$
eV. However, another two peaks of $t_2$ DOS are clearly seen just
at $\varepsilon_F$ and $0.5$ eV below it. The origin of these
peaks becomes more clear when the Cu $t_2$ DOS is compared to the
density of $p$ states of the P$_1$ ion closest to Cu. The latter
shows two prominent peaks exactly at the same energies. Similar
peaks can also be observed in Ga$_1$ DOS as well as in DOS of the more
distant P and Ga ions not shown in Fig.~\ref{Fig:dos32}. An
analysis of the partial occupations shows that of 2 holes ($h$)
created by the Cu impurity only 0.18 $h$ is provided by the Cu $t_2$
states. Another 0.48 $h$ is distributed over the $p$ states of 4
$P_1$ ions whereas the remaining 1.36 $h$ is spread over more
distant neighbors.

It is worth noting that in spite of the appearance of the narrow
DOS peak exactly at $\varepsilon_F$ the spin-polarized calculations
failed to produce a ferromagnetic solution even for the highest Cu
concentrations studied. Apparently, this can be explained by the
delocalized character of the states responsible for the peak and an
insufficient strength of the Hund's exchange coupling for P and Ga
$p$ states, which give the dominant contribution to the
corresponding bands. At the same time, the contribution of Cu $d$
states, for which a strong on-site exchange interaction is expected,
to the peak at $\varepsilon_F$ is relatively small.

We also performed test calculations for a few values of $x$ in
$E_{x}$Ga$_{1-x}$P, in which a Ga ion was substituted by a vacancy
$E$. A vacancy creates one more hole in the valence band as
compared to Cu. Nevertheless, in the vicinity of the Fermi level
the band structures calculated for $E_{x}$Ga$_{1 - x}$P are
similar to those for Cu-doped GaP. In particular, the density of
P$_1$ states at and just below $\varepsilon_F$ has the same
two-peak shape. These peaks are also reflected in the density of
$E$ $d$ states of the $t_2$ symmetry, however, they are much less
pronounced than the corresponding peaks of Cu $t_2$ DOS.
Significantly higher peaks can be observed in the density of $E$
$p$ states which also transform according to the $t_2$
representation.

Thus, we may conclude that the bands crossing $\varepsilon_F$ in
Cu$_{x}$Ga$_{1 - x}$P are mainly formed by the $p$ states of the
nearest to the Cu impurity P$_1$ ions that split off from the top
of the GaP valence band as a result of breaking of the covalent P
$p$ -- Ga bonds at the impurity site. These states have $t_2$
symmetry and hybridize strongly with the corresponding Cu $d$
states. These states are, however, rather extended, which leads to
a relatively strong dispersion of the split-off bands even for
$x$=0.016.

\subsection{Cu-related energy states of isolated impurity}

Before turning to the calculation of the Cu-impurity related levels in
the host GaP, let us look at the energy dependence of the self
energy part (\ref{massoper}), which is responsible for the
renormalization of $3d$ levels due to the hybridization with the host
band states. The hybridization matrix elements $M_{n{\bf k},i_a}$
are calculated by means of Eq. (\ref{matelem}) using the Cu 3$d$
impurity wave functions and LAPW functions of the GaP host. Since
the LAPW wave functions are defined within the volume subdivided
into two muffin-tin parts and the surrounding volume, the
integration in Eq. (\ref{matelem}) is carried out in all three
parts separately accounting for all the hybridization
contributions as well as for the covalency induced non-spherical
components of the difference potential.

\begin{figure}[tbp!]
\begin{center}
\includegraphics[width=0.95\columnwidth]{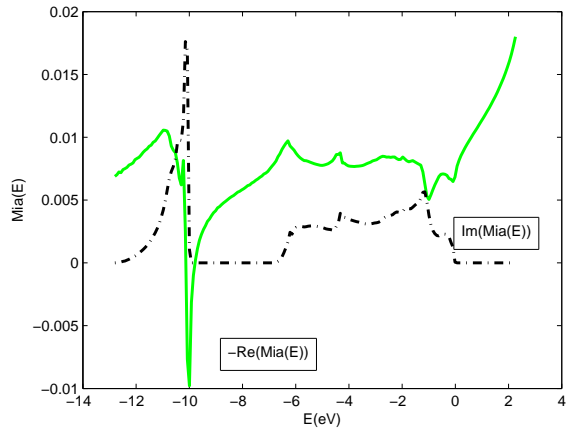}
\end{center}
\caption{(Color online)Real and imaginary parts of mass operator
${\cal M}_{i_a}(\varepsilon)$ for (Ga,Cu)P.} \label{M}
\end{figure}

Figure \ref{M} represents the real and imaginary parts of ${\cal
M}_{i}(\varepsilon)$ obtained for the (Ga,Cu)P compound. Here
index $i$ represents one of the components of the $t_2$
irreducible representation. Comparison of ${\rm Im} {\cal
M}_{t_2}(\varepsilon)$ with the density of band states which is
shown as ${\rm Im}(G_h^0)$ in Fig. \ref{Gh} demonstrates that the
weighting of the density of states with the squared hybridization matrix element reproduces the general shape and van Hove singularities of the partial $p$-component of DOS. The differences between ${\rm Re}(G_h^0)$ and ${\rm Re}{\cal
M}_{t_2}(\varepsilon)$ are more noticeable. Both these functions
are sums of the Hilbert transforms of the DOS and weighted DOS for
all the valence and conduction bands, respectively. Therefore
these function not only map the singularities of DOS in the given
band on the singularities of its Hilbert transform but also
accumulate asymptotic contributions of higher and lower bands at the
given $\varepsilon$. This accumulation results in a noticeable
smoothing of the ${\cal M}_{t_2}(\varepsilon)$ function in the $
-6$ to 0 eV range. Besides, weighting with $M^2_{t_2}(\varepsilon)$ strongly reduces the amplitude of ${\rm Re}{\cal M}_{t_2}(\varepsilon)$ in comparison with ${\rm Re}(G_h^0)$. Such strong reduction means that the
hybridization-induced renormalization of the atomic $3d$ levels of
the isolated Cu impurity is small enough, and their positions are
predetermined mainly by the impurity core potential and Coulomb
interaction within the muffin-tin sphere $r_D$.

To compare the energy spectrum of the Cu impurity in GaP obtained
by the Green function method with that given by LDA in the
supercell calculation scheme, we first compute this spectrum by
solving Eq. (\ref{resontscat}) within the LDA scheme without the
second term $\Delta V_{ii}^U$ in the impurity potential
(\ref{dev}). Both the resonant and short range potential
components of impurity scattering were taken into account. These
calculations yield the value $\varepsilon_v-0.66$ eV for the
impurity $d_{t_2}$ resonance in the valence band, which is higher
than that in the supercell calculation, and the $d_e$ peak lies
slightly above this level. Apparently, these peaks are related to
the van Hove singularities in the heavy hole band. These
resonances are shallower than those seen in the supercell DOS
(Fig. \ref{Fig:dos32}). As was mentioned above, the $d_e$ peak in
the latter structure is located at $\varepsilon_v - 2.5$ eV.
However, one should remember that the center of gravity of the
valence band DOS is shifted downward with respect to its position
in the pure GaP due to the transformation of $d_e$ and $d_{t_2}$
levels into $d$-bands (see Fig. \ref{Fig:bnd32}). Potential
scattering built in the self energy part $\widetilde M_i(z)$ in
Eq. (\ref{resontscat}) results in the appearance of an empty
impurity level at $\varepsilon_v + 0.168$ eV. This acceptor level
may be identified with the $x \to 0$ limit for the P related
$p$-structure at the top of the valence band in the supercell DOS
(Fig. \ref{Fig:dos32}, middle panel). The occupation of the
impurity $d$-shell in this case is close to 10, like in the
supercell calculations.

The computation of the impurity spectrum within the LDA+U scheme
yields a self-consistent solution for the electron spectrum only
for the transition state 3d$^{8.5}$ of Cu impurity. This solution
is described below. First, we determined the position of non-perturbed 3d-level of the Cu atom and the correlation parameters $U - J$. The isolated
impurity energy $\varepsilon_{i}(+8.5) = - 20.9$ eV is calculated
by means of the semi-relativistic RATOM program\cite{atom} for the
3d$^{8.5}$ configuration, which corresponds to the concept of the
transition state adopted in this paper. The intraatomic Coulomb
repulsion of the $d$-electrons is treated in the LDA + U
approximation and $m$-dependent Coulomb integrals (\ref{slat}) are
calculated. The choice of the parameters $U = 4.5$ eV and $J = 0.7$
eV is based on the analysis of the occupation numbers in the
transition state approach.\cite{Sato} The self-consistent single
electron 3$d$-level for the embedded Cu impurity in the 3d$^{8.5}$
configuration is in resonance with the valence band of GaP host
crystal, and the impurity-related resonance and discrete states
are found as solutions of Eq. (\ref{resontscat}).

\begin{widetext}

\begin{figure}[tbp!]
\begin{center}
\includegraphics[width=0.95\columnwidth]{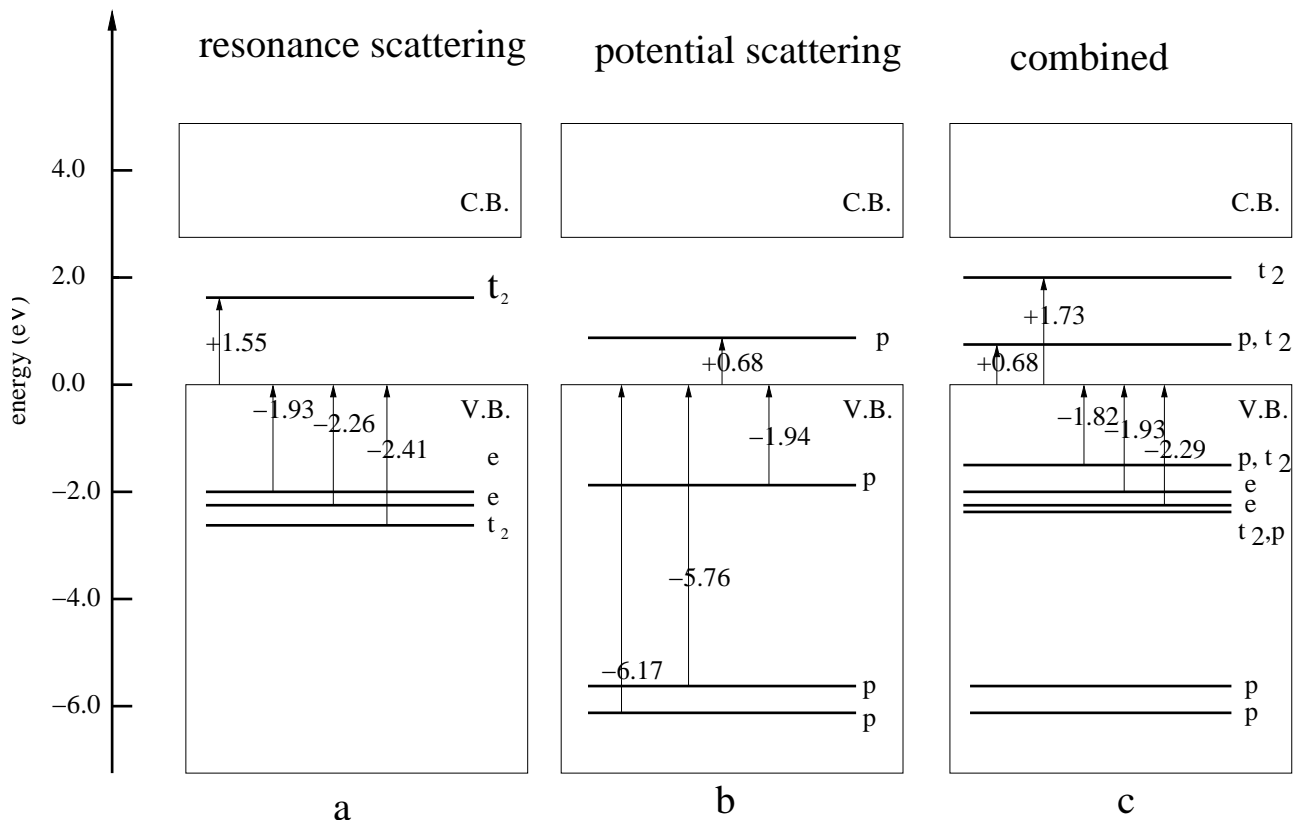}
\end{center}
\caption{ Electronic structure of (Ga,Cu)P, calculated from Eq.
(\ref{resontscat}). The lowest of the five levels in the left
panel correspond to the states $(2,\pm1)$, the next $(l,m)$ levels
are classified as (2,-2),(2,0) and (2,+2) (bottom-up). See the
text for further discussion} \label{elstr}
\end{figure}

\end{widetext}

Figure \ref{elstr} depicts the  electronic structure of (Ga,Cu)P
calculated by the Green function method. We present here three versions of the
calculations which account for: (a) resonant scattering, (b) short
range potential scattering, and (c) combined case.

In the resonant scattering approximation (Fig. \ref{elstr}a) where
the term $Q^{-1}(z)$ is omitted in Eq. (\ref{resontscat}), there
are four occupied levels in the valence band and one empty level
in the energy gap. The occupied states correspond to the
configuration Cu($d^8)$ of the impurity ion. These levels reflect
the multiplet structure of this configuration. Although we used
the orbital quantum numbers in our computation procedure, the
calculated electron density distribution reveals the $T_d$ point
symmetry of the impurity surrounding. In terms of the
corresponding cubic harmonics the lowest state has the $t_2$
symmetry, the two next levels belong to the $e$-representation,
and the empty state in the energy gap is the $t_2$ state of the
configuration $d^9$. In terms of the Allen diagrams (\ref{allen})
these levels correspond to the addition energy
$\varepsilon_{t_2}^{9/8} = E(e^4t_2^5) - E(e^4t_2^4)$ and
$\varepsilon_e^{9/8} = E(e^4t_2^5) - E(e^3t_2^5)$ for $d_{t_2}$
and $d_e$ quantum numbers, respectively (see similar
classification for (Ga,Ni)P in Ref. \onlinecite{Zung85}). The
final $3d^8$ states belong to the $^3T_2(F)$ and $^3T_1(F)$
representations in the Tanabe-Sugano
classification.\cite{Zung86a,KF94,FK86,Zung85}

The energy interval between the multiplet of occupied levels in
the valence band and the empty level in the energy gap is
$\lesssim 4$eV, which is comparable with the value of the input
parameter $U - J = 3.8$ eV. The hybridization renormalization due to
the self energy ${\cal M}(z)$ in Eq. (\ref{resontscat}) is 0.115
eV for the occupied levels and 0.182 eV for the empty level. In
the calculation procedure described above, the difference in
hybridization shifts for $t_2$ and $e$-levels was neglected,
because the hybridization (ligand field) contribution is small
enough for the Cu impurity ion.

Figure \ref{elstr}b exhibits the net contribution of potential
scattering (\ref{dev1}) to the formation of impurity-related
states. The levels shown in this figure are obtained from
(\ref{resontscat}) with ${\widetilde{\cal M}}$ substituted for
$Q^{-1}$ [see Eq. (\ref{selfmod})]. The states in the occupied
part of the spectrum are the impurity resonances in the valence
bands around the maxima of the partial $p$-wave contributions at the
energies $\sim -6$eV and $\sim -2$eV (cf. Figs. \ref{Fig:dos32}
and \ref{M}). The $p$-level arises at the energy $+0.68$eV above
the top of the valence band.

Both the $d$- and $p$-like states are found in the solution of Eq.
(\ref{resontscat}) with the full self energy ${\widetilde{\cal
M}}$ (Fig. \ref{M}c). The most significant difference between the
combined spectrum of Fig. \ref{M}c and those of Fig. \ref{M}a,b is
the noticeable hybridization between $p$ and $d_{t_2}$ resonances
in the valence band, whereas thee $d_e$ levels are only slightly
shifted. The shallow $p$-level in the energy gap is pinned to its
original position shown in Fig. \ref{M}b, in spite of $dp$
hybridization. All these results agree with qualitative
predictions of the analytical model taking into account both
resonant and short-range potential stattering.\cite{FK86}

There is no straightforward way to compare the results of LDA+U
calculations with those obtained within the LDA scheme, since the
former method uses the fitting parameters $U,\ J$, whereas the
latter one is based on the variational approach, which formally
gives the solution corresponding to the minimal total energy. We
only may estimate the total energies of the two solutions by
comparing the positions of the impurity levels obtained by both
methods within the same Green function approach. LDA procedure
gives the occupied $e$ and $t_2$ levels at the energies $\sim
\varepsilon_v - 0.64$ to 0.66 eV below the top of the valence band
and the shallow $p$-level at the energy $\varepsilon_v + 0.168$
eV, which corresponds to the configuration $d^{10}\bar p^2$: two
holes neutralize the excess charge in the $d$ shell, which means
that the triply degenerate $p$-level is occupied by one electron.
In the LDA+U solution the occupied $t_2$ and $e$-levels lie
essentially deeper in the valence band at the energies $\sim
\varepsilon_v - 2.3$ to 1.8 eV, Cu ion behaves as the
isoelectronic impurity ${\rm Cu}^{3+}(d^8)$, and the acceptor
$p$-levels are triply occupied in the neutral impurity state. The
comparison of single-electron energies for the two solutions gives
the energy gain $\sim 10.7$ eV for the latter state. It is hardly
probable that the exchange-correlation contribution may change the
energy balance in favor of a state with the fully occupied 3$d$
shell of the Cu impurity.

Comparing the electronic structures of (Ga,Cu)P obtained by the
supercell and Green function methods, one may indicate both similarities and
dissimilarities in the description of impurity-related states.

First, both methods provide the same mechanism for formation of the
shallow $p$-levels in the energy gap of the host material, which merge
into the impurity band at a high enough dopant concentration. These
levels are split off from the top of the valence band and partially
hybridized with the $t_2$ levels in the valence band.

Second, the spectral density of the impurity related
$d_\gamma$-states is concentrated mainly in the valence band with
the $d_{t_2}$ component lying below the $d_e$ component. Here,
however, the important difference between the two approaches
should be emphasized. As was mentioned above, the $d_\gamma$
resonances calculated within the Green function LDA approximation
are shallower than those found by means of the supercell approach.
One may see also a difference in the $d_e - d_{t_2}$ splitting: it
can be estimated as $\sim 0.5$eV in the supercell calculations and
as $\lesssim 0.3$eV in the Green function calculations. The main
reason of this difference is the fact that the impurity
$3d$-levels are transformed in an effective $d$-bands in the
periodic supercell structure, and the hybridization repulsion
between the two Bloch waves is stronger than that between the
localized $d$-levels and periodic partial $p$-waves in the Green
function approach. The same argument is valid for the quasiband
method used in the calculations of Ref. \onlinecite{Zung85}, where
the Cu-related $d$ - levels are located even deeper than in our
supercell calculations at the energies $\sim 3-4.5$ eV below the
top of the valence band of GaP.

The most important difference between the results described in
Subsections IVA and IVB is of course the difference in the
electron configuration of Cu impurity, which is $d^{10}\bar p^2$
in the supercell calculations and $d^8$ in the Green function calculations.
Available experimental data \cite{Gupta} are in favor of the
configuration $d^9\bar p$. At this stage we have no exhaustive
explanation of these discrepancies. First, our scheme should be
extended to the spin-unrestricted LSDA solution and to the
multiimpurity case. We expect that the charge configuration of Cu
ions is highly sensitive both to the spin state and to the
interimpurity coupling. Second, more experimental investigations
are necessary, which would reveal the role of concomitant defects,
the annealing conditions, the thickness of the film and other
technological factors. It is also worthwhile checking whether the
use of LDA+U method in the supercell approach may result in the
configuration with an incomplete $3d$ shell of the Cu impurity. We
leave all these questions for further investigations.

\section{Concluding remarks}

The numerical solution of the Dyson equation (\ref{resontscat})
derived by means of the Green function method reveals similarities and
dissimilarities between the electronic structures of the Mn impurity
(half-filled 3$d$ shell in atomic state) and Cu impurity (completely filled 3d shell in atomic state) substituting for Ga in zinc blende semiconductor. Our
calculations show that unlike Mn, which retains its stable
half-filled 3d$^5$ shell in the host GaAs and GaP
crystals,\cite{McD,merge} the Cu impurity may release some of its
$d$-electrons from the stable filled shell 3d$^{10}$  to minimize the total
energy of doped crystal, at least
in the wide-gap GaP. Our theoretical result partially agrees with
the experimental observation of Cu ions with unfilled 3d shell in
GaP.\cite{Gupta} It paves a way to theoretical explanation of the ferromagnetic ordering in Ga$_{1-x}$Cu$_x$P crystals, although for this purpose further development of the Green function method is necessary. The results of the numerical study of magnetic ordering by means of the Green function method will be published
elsewhere.

This work is partially supported by the Max-Planck Gesellschaft
during the stay of O.F., K.K. and V.F. in MPIPKS (Dresden), where
this work was completed.

\appendix
\section{Details of computational scheme}

In order to realize the GF approach in a computational scheme we
make use of the local density approach (LDA)\cite{Far} and its
LDA+U modification \cite{Anis93} which  accounts for a strong
electron-electron interaction. The approximation \cite{Cap} is used for the exchange-correlation potential. The band structure of the GaP semiconductor is calculated by means of the ab-initio full potential all electrons LAPW method.\cite{FaVa} This method presents the charge density and the crystal potential as a series of the spherical harmonics inside the muffin-tin
spheres and of the plane waves outside the spheres. The
self-consistent electronic band structure is determined by solving
a single particle Dirac equation by using the variational method
in LAPW - function basis \{$\Psi^{LAPW}_{n{\bf k}}({\bf r})$\}. In
order to evaluate the Coulomb part of the crystal potential we use
the concept of multipole potentials and solve the Dirichlet
problem for the sphere with all the contributions being treated on
equal footing.\cite{FaVa} The exchange-correlation potential is
approximated by the Pad\'e approximant technique in order to
interpolate accurately the recent Monte Carlo results with the RPA
spin-dependent data.\cite{Cap} The Fourier components of the
exchange - correlation potential in the interstitial region are
fitted in the least square method by applying the singular value
decomposition procedure. The charge density in the
interstitial region is calculated in ca. 2000 to 3000 random
points in the irreducible wedge of the Wigner-Seitz cell.

In order to find the self energy ${\cal M}_{i}(z)$, one has to
calculate the matrix elements $M_{n{\bf k},i}$ between the band
states $|n{\bf k}\rangle$ and the states $|i\rangle\equiv
|plm\rangle$ of the impurity atom. A computational scheme based on
the augmented Green functions\cite{KZ}
\begin{equation}\label{augmgreen}
 {\sf{G}}^0(z) = \left(
\begin{array}{cc}
G^0_A(z) & 0 \\
&\\
0 & G^0_h(z)
\end{array}
\right)
\end{equation}
is developed for this sake. Here
$$
G^0_A({\bf r},{\bf r}';z) =\sum_{i=p,l,m} \frac{\phi_i({\bf
r})\phi^*_i({\bf r})}{z - \varepsilon_{i}}
$$
is the impurity Green function, whereas the host crystal is
represented by
$$
G_h^0({\bf r},{\bf r}'; z) = \sum_{n = 1}^P \sum_{{\bf k} \in IBZ}
\frac{\Psi^{LAPW}_{n{\bf k}}({\bf r}) {\Psi^{LAPW}_{n{\bf
k}}}^*({\bf r}')}{z - \varepsilon_{n{\bf k}}}.
$$

The wave functions of electrons localized in the impurity 3$d$ shell
are defined within the impurity sphere $r \leq r_D$ (see Fig. \ref{spheres}):
$\phi_{plm}({\bf r}) = R_{pl}(r) Y_{lm}(\widehat{{\bf r}})$. The
radial parts of these functions are defined as solutions of the
equation
\begin{equation}\label{phid}
\left[- \nabla^2 + V_h^0(r) + \Delta V(r)\right]R_{pl}(r) =
\varepsilon_{pl} R_{pl}(r),
\end{equation}
and the angular parts are represented by the spherical harmonics. The
Bloch wave functions are expanded in the reciprocal wave vectors ${\bf
k}_\varrho = {\bf k} + {\cal K}_\varrho$
$$
\Psi^{LAPW}_{n{\bf k}}({\bf r}) = \sum_{\varrho = 1}^N v_n({\bf
k}_\varrho) \varphi_{{\bf k}_\varrho}({\bf r})
$$
where
\begin{widetext}
$$
\varphi_{{\bf k}_\varrho}({\bf r}) =
\frac{1}{\sqrt{\Omega_0}}\left[ e^{i{\bf k}_\varrho {\bf r}}
\Theta_{int}({\bf r}) + \sum_s \Theta_s( \mbox{\boldmath$\rho$})
e^{i{\bf k}_\varrho{\mbox{\boldmath$\tau$}}_s}4\pi r_s^2\sum_{lm}
i^l \Phi^{(s)}_{lm}({\bf k}_\varrho, \mbox{\boldmath$\rho$})
Y^*_{lm}(\widehat{k}_\varrho) \right]
$$
\end{widetext}
with
$$
\Phi^{(s)}_{lm}({\bf k}_\varrho, \mbox{\boldmath$\rho$}) =
$$$$
\left[a_l^{(s)} ({\bf k}_\varrho) {\cal{R}}^{(s)}_l(\varepsilon_l,
\rho) + b_l^{(s)} ({\bf k}_\varrho)
\dot{{\cal{R}}}^{(s)}_l(\varepsilon_l, \rho)
\right] Y_{lm}(\widehat{\mbox{\boldmath$\rho$}}).
$$
The following notations has been used above:
$$
\Theta_{int}({\bf r}) = \left\{
\begin{array}{ll}
1, & {\bf r} \in \Omega_{int}\ -\ \mbox{volume of the}\\
&\ \mbox{interstitial
region}\\
&\\
0, &\ \mbox{otherwise},
\end{array} \right.
$$
$$
\Theta_s(\mbox{\boldmath$\rho$}) = \Theta_s({\bf r} - {\mbox{\boldmath$\tau$}}_s)
= \left\{
\begin{array}{ll}
1, & {\boldmath \rho} \in \Omega_{s}\ -\ \mbox{volume of the} \\
& \mbox{$s$ sphere
region}\\
&\\
0, &\ \mbox{otherwise},
\end{array} \right.
$$
$v_n({\bf k}_\varrho)$ are eigenvectors of LAPW variation
procedure; $n$ is the number of the accounted energy bands,
$\Omega_0$ is the volume of the Wigner-Seitz cell, $a_l^{(s)}
({\bf k}_\varrho)$ and $b_l^{(s)} ({\bf k}_\varrho)$ are the
muffin-tin coefficients in the LAPW method, and
$\dot{{\cal{R}}}^{(s)}_l(\varepsilon_l,\rho) =
\frac{\partial}{\partial\varepsilon}{{\cal{R}}}^{(s)}_l(\varepsilon,
\rho)|_{\varepsilon_l}$ for the fixed energy
$\varepsilon_l$. ${\cal{R}}^{(s)}_l(\varepsilon_l,
\rho)$ is the radial part of the LAPW function.

\subsection{Choice of the localized basis}

Calculations of the electronic structure of defects in crystals
are usually based on the pseudopotential or LCAO + pseudopotential
approach\cite{Bernholc,Baraff}. This method requires a large
number of the Gaussian orbitals and calculation of their overlap
integrals. Instead we perform here an all-electron calculation
which allows one to realize the spin-polarization LDA + U scheme.
This approach uses the basic set of  $N_D$ functions
\begin{equation}\label{wf_1}
\chi_\mu ({\bf r}) \equiv \chi_{pLM}(r,\vartheta,\varphi) =
$$$$
\left\{
\begin{array}{cc}
F_{{\bf p}L}(r) Y_{LM}(\hat
{\bf r}), & \mbox{for}\ r\leq r_D,\\ &\\
j_L(\kappa_{{\bf p}L}r) Y_{LM}(\hat {\bf r}) & \mbox{for}\ r>r_D
\end{array}
\right.
\end{equation}
Here  $L$ is a non-negative quantum number and $-L \leq M \leq L$,
the inverse length $\kappa_{pL}$ is defined by zeros of the Bessel
function $j_L(\kappa_{L_p}r_{emb}) = 0$ for the radius $r_{emb}$
of the embedded sphere; $p$ is the integer number enumerating
these zeros. The radial part of the wave function (\ref{wf_1}) is
\begin{equation}\label{radial}
F_{pL}(r) = a_L(\kappa_{pL}) R_L(\varepsilon_L,r) +
b_L(\kappa_{pL}) \dot R_L(\varepsilon_L,r).
\end{equation}
Here the parameters
$$
a_L(\kappa_{pL}) = \frac{\dot R_L j_L'(\kappa_{pL}r_D) - {\dot
R_L}' j_L(\kappa_{pL}r_D)}{\dot R_L R_L' - R_L \dot R_L'}
$$
$$
b_L(\kappa_{pL}) = \frac{R_L' j_L(\kappa_{pL}r_D) - R_L'
j_L'(\kappa_{pL}r_D)}{\dot R_L R_L' - R_L \dot R_L'}
$$
are used to match the function (\ref{radial}) to the Bessel
functions outside the muffin-tin region, $R_L\equiv
R_L(\varepsilon_L,r)$, $R_L' = \displaystyle\left.\frac{d
R_L(\varepsilon_L,r)}{dr}\right|_{r=r_D}$,
$\displaystyle\left.\dot R_L = \frac{d
R_L(\varepsilon_L,r)}{d\varepsilon_L}\right|_{r=r_D}$.

The above basis $\chi_\nu({\bf r})$ was used in the Cholesky
decomposition ${\textsf S}= {\textsf L}\cdot {\textsf L}^\dagger$
for the overlap matrix
$$
S_{\mu\nu} = \int_{\Omega_{emb}} \chi_\mu^*({\bf r}) \chi_\nu({\bf
r}) d{\bf r}
$$
in order to obtain the orthonormal basis
$$
\widetilde\chi_\mu({\bf r}) = \sum_{\mu'} ({\textsf
L}^{-1})^\dagger_{\mu\mu'} \chi_{\mu'}({\bf r}).
$$

Then the Green function of the host crystal is projected onto the
localized basis
$$
G_{h,\mu\nu}^0(z) = \sum_{n=1}^M \sum_{{\bf k} \in IBZ}
\frac{\langle \widetilde\chi_\mu|\Theta|\Psi^{LAPW}_{n\bf k}\rangle
\langle \Psi^{LAPW}_{n\bf k}|\Theta|\widetilde\chi_\nu\rangle}{z -
\varepsilon_{n\bf k}}
$$
and calculated by means of the analytical tetrahedron method
\cite{tet} within the  irreducible part of the Brillouin zone
({\it IBZ})
$$
G_{h,\mu\nu}^0(z) = \sum_{n=1}^M \sum_{{\bf k} \in IBZ} w_{n\bf k}(z)
{\cal F}_{n{\bf k},\mu\nu}.
$$
Here
$$
{\cal F}_{n{\bf k},\mu\nu} = \langle
\widetilde\chi_\mu|\Theta|\Psi^{LAPW}_{n{\bf k}}\rangle \langle
\Psi^{LAPW}_{n{\bf k}} |\Theta|\widetilde\chi_\mu\rangle.
$$
The coefficients $w_{n\bf k}(z)$ depend only on the dispersion
relation $\varepsilon_{n{\bf k}}$ and can be computed only once.

\subsection{Self energies for impurity Green function}

The impurity Green function (\ref{dyson}) contains several self
energy corrections to the atomic  levels $\varepsilon_i$. Two of
them given by Eqs. (\ref{dev2}),(\ref{dev3}) arising from the Coulomb
interaction are responsible for the multiplet structure of the energy
levels, potential contribution (\ref{dev1}) results in the crystal
field splitting of these levels, and the resonance self energy
(\ref{selfmod}) is the analog of ligand field correction in conventional
theory of transition metal impurities.\cite{FK86}
This section discusses the calculation of the two last terms within the Green function formalism.

The resolvent operator ${\sf \Delta G}(z)$ and the corresponding
density variation $\Delta\rho({\bf r})$ is  calculated both for
the host block [$\Delta G(z)$, $\Delta\rho({\bf r})$] and for
impurity block [${\sf \Delta G}_{ii}(z)$, $\Delta\rho_{i}({\bf
r})$] of the secular matrix (\ref{augmgreen}). When calculating
the contour integrals resulting in (\ref{deltarho_1}) we use
semi-circular contour from the bottom of the valence band
$\varepsilon_b$ to the Fermi energy $\varepsilon_F$. The charge
dependent difference potential $\Delta V({\bf r})$ is  not
necessarily spherically symmetric. We define the substitution
impurity potential as the difference
\begin{equation}\label{sub-1}
\Delta V[\rho({\bf r})] = V_{eff}[\rho({\bf r})] -
V_h^0[\rho_h^0({\bf r})]
\end{equation}
between the true self-consistent effective potential
$V_{eff}[\rho({\bf r})]$ and the effective self-consistent
potential $V_h^0[\rho_h^0({\bf r})]$ of the host crystal, both
taken in the LDA approximation. Here $\rho({\bf r})$ and
$\rho_h^0({\bf r})$ are the respective electron densities.

The impurity correction to the host Green function of the crystal
induced by the potential (\ref{dev1})
$$
\Delta{\textsf{G}}(z) ={\textsf{G}}(z) - {\textsf{G}}_h^0(z)
$$
is found from the corresponding Dyson equation\cite{Wachutka}
$$
\Delta\textsf{G}(z) = \left[\left({\textsf I} - \widetilde {
{\textsf G}}^0_h(z)\cdot \Delta{\cal V} \cdot ({\textsf
L}^{-1})^\dagger \right)^{-1} - {\textsf I} \right] \widetilde{
\textsf G}_h^0(z).
$$
Here
$$
(\Delta{\cal V})_{\mu\nu} = \int_{\Omega_{emb}} \chi^*_\mu({\bf r})
\Delta V[\rho({\bf r})] \chi_\nu({\bf r}) d{\bf r}
$$
and ${\textsf I}$ is a unit matrix.

The density variation is calculated using the equation
\begin{equation}\label{deltarho}
\Delta \rho({\bf r}) = \mbox{Im} \sum_{\mu=1}^{N_D}
\sum_{\nu=1}^{N_D} \widetilde{\Delta\rho}_{\mu\nu} \chi_\mu({\bf
r}) \chi^*_\nu({\bf r})
\end{equation}
where
$$
\widetilde{\Delta\rho}_{\mu\nu} = \left(({\textsf L}^{-1})^\dagger
{\Delta\rho} {\textsf L}^{-1}\right)_{\mu\nu}
$$
and
\begin{equation}\label{deltarho_1}
\Delta\rho = - \frac{1}{\pi} \int_{\varepsilon_b}^{\varepsilon_F}
\Delta\textsf{G}(z) dz.
\end{equation}
The lower integration limit $\varepsilon_b$ is chosen to include
all the relevant band and impurity states, $\varepsilon_F$ is the
Fermi energy. To compute the integral (\ref{deltarho_1}), we
introduce the contour $C$ in the complex plane $z$ enclosing all
the poles up of the Green function up to the Fermi energy in the charge density integration.

With $\Delta \rho({\bf r})$ calculated by means of Eqs.
(\ref{deltarho}) and (\ref{deltarho_1}) we calculate anew the
charge dependent impurity potential
\begin{equation}
\Delta V({\bf r}) = \sum_{LM} \Delta V_{LM}(r) Y_{LM}(\widehat{\bf
r})
\end{equation}
in the "embedded cavity", which is not spherically symmetric. The
density variation can be similarly represented as
\begin{equation}
\Delta\rho({\bf r}) = \sum_{LM} \Delta \rho_{LM}(r)
Y_{LM}(\widehat{\bf r})
\end{equation}
where one readily obtains
\begin{equation}
\Delta\rho_{L''M''}(r) =
$$$$
\sum_{pp'}\sum_{LL'} \widetilde{\Delta
\rho}_{pL,p'L'} \Gamma_{pp'}(L,L';r) \sum_{M=-L}^L G_{L'M'L''M''}^{L,M'
+ M''}
\end{equation}
with
\begin{widetext}
\begin{equation}
\Gamma_{pp'}(L,L';r) =
$$$$
\left\{
\begin{array}{ll}
a_L({\kappa_{pL}}) a_{L'}(\kappa_{p'L}) R_L(\varepsilon_L,r)
R_{L'}(\varepsilon_L,r) + a_L(\kappa_{pL}) b_{L'}(\kappa_{pL})
R_L(\varepsilon_L,r) \dot R_{L'}(\varepsilon_L,r) + & \\
a_{L'}(\kappa_{pL}) b_L(\kappa_{pL}) \dot R_L(\varepsilon_L,r)
R_{L'}(\varepsilon_{L'},r) + b_L(\kappa_{pL}) b_{L'}(\kappa_{p'L'})
\dot R_L(\varepsilon_L,r) \dot R_{L'}(\varepsilon_L,r), &
\mbox{for}\ 0< r \leq r_D\\
&\\
j_L(\kappa_{pL}r) j_{L'}(\kappa_{p' L'}r)& \mbox{for}\ r_D < r
\leq r_{emb}
\end{array}
\right.
\end{equation}
\nonumber
\end{widetext}
and
$$
G^{L' M+M''}_{LML''M''} = \int_S dS Y^*_{L'' M''}(\vartheta,\varphi)
Y_{L M}(\vartheta,\varphi) Y^*_{L' M'}(\vartheta,\varphi)
$$
being the Gaunt coefficients.

Next we separate the impurity and host parts in the density
correction
$$
\widetilde{\Delta\rho}_{L''M''}(r) = \Delta\rho_{L''M''}(r) +
\Delta\rho^{(s)}_{L''M''}(r)
$$
where
\begin{eqnarray}
&&\Delta\rho_{L''M''}(r) = \sum_{\mu = 1}^{N_D}
\Delta\rho_{\mu\mu}
\Gamma_{pp}(L,L; r) G^{LM}_{L M L'' M''} \nonumber \\
&&+ 2 \sum_{\mu = 2}^{N^D} \sum_{\mu' = 1}^{\mu - 1} \mbox{Re}
(\Delta\rho_{\mu\mu}) \Gamma_{pp'}(L,L';r) G^{L'M'}_{L M L'' M''}
\nonumber
\end{eqnarray}
is the host contribution, and
$$
\Delta\rho^{(s)}_{L''M''}(r) = R^2_{pl}(r) \sum_{m = - l}^{l}
\Delta \rho_{lm,lm} G_{lmL'' M''}^{lm + M''}
$$
is the substitution impurity contribution. The functions
$R_{pl}(r)$ are the radial parts of the impurity centered local
orbitals (\ref{phid})

Using this density correction, we calculate the impurity related
self energy $\Delta V_{ii}^{LDA}$
\begin{equation}
\Delta V _{ii}^{LDA} =
$$$$
\sum_{L''M''} G_{lm - M''L''M''}^{lm} \int_0^{r_{emb}}dr\ r^2
\Delta V_{L''M''}(r) {R}^2_{pl}(r)\nonumber
\end{equation}
and substitute it into the Green function (\ref{LDA+U}).

Self energy correction ${\cal M}_i$ in (\ref{selfmod}) contains
the matrix elements (\ref{matelem}). After substituting the
potential (\ref{sub-1}) in the integral (\ref{matelem}) and matching
the boundary conditions in accordance with the procedure described
above, the hybridization matrix element acquires the form
(\ref{matelspher}).

\end{document}